\documentstyle[preprint,aps,eqsecnum]{revtex}

\tighten

\begin{document}

\draft

\preprint{CRN 95-11}

\title  {
                     Time-odd components in the mean field of rotating
                                  superdeformed nuclei
        }
\author {
                       J.Dobaczewski$^{(1,2)}$ and J. Dudek$^{(1)}$
        }
\address{
              $^{(1)}$Centre de Recherches Nucl\'eaires,
                      IN$_2$P$_3$--CNRS/Universit\'e Louis Pasteur        \\
                      F-67037 Strasbourg Cedex 2, France                  \\
              $^{(2)}$Institute of Theoretical Physics, Warsaw University \\
                      ul. Ho\.za 69, PL-00681 Warsaw, Poland
        }

\maketitle

\begin{abstract}
Rotation-induced time-odd components in the nuclear mean field
are analyzed using the Hartree-Fock cranking approach with
effective interactions SIII, SkM*, and SkP. Identical
dynamical moments ${{\cal J}^{(2)}}$ are obtained for pairs of
superdeformed bands
$^{151}$Tb(2)--$^{152}$Dy(1) and $^{150}$Gd(2)--$^{151}$Tb(1).
The corresponding relative alignments strongly depend on which
time-odd mean-field terms are taken into account in the
Hartree-Fock equations.
\end{abstract}

\pacs{PACS number(s): 21.60.-n, 21.60.Ev, 21.60.Jz}

\narrowtext

\section{Introduction}
\label{sec1}

A description of ground and excited states in terms of a mean
field is a well established approach in nuclear structure.  Most
of the nuclear phenomena can be considered as manifestations of
the mean field properties, or the mean field can be used as a
suitable framework and the first-order approximation.
The mean field
results from averaging the nucleon-nucleon interactions
over states of individual
nucleons. The averaging procedure, which must take into account
the Fermi statistics of nucleons, can be formalized in terms of
a variational approach, and leads to the well-known Hartree-Fock
(HF) selfconsistent equations \cite{RS80}.

Dynamic or time-dependent phenomena can be described by a
corresponding time-dependent HF method. The nuclear state is
then represented by a one-body density matrix, which evolves
in time according to the Hamilton equations, and represents a
motion of a wave packet. Such an approach has been applied to
genuinely time-dependent problems, like nuclear reactions, but it
is in fact best suited to describe stationary collective states.

The nuclear rotation is an example of a collective motion for
which linear combinations of
stationary states of a given spin can be identified
with a rotating wave packet.  In this case, a transition from
the time-dependent HF theory to a stationary problem can be
achieved by introducing rotating intrinsic frame of reference.
In this frame, the equations of motion are time-independent,
however, the resulting density matrix is not invariant with
respect to the time-reversal operator. As a consequence, the
mean field obtained for such a density matrix also loses its
time invariance, and acquires new terms which are odd with respect
to the time reversal.

Properties of nuclear time-even mean fields are known rather
well, because they are reflected in multiple static phenomena
which can be studied experimentally \cite{AFN90a}.  On the other
hand, very little is known about properties of the time-odd mean
fields. Most studies were up to now carried out either within
the adiabatic \cite{BV78} or semiclassical approximations
\cite{GV79}.  In particular, the adiabatic approximation to the
nuclear translation, rotation, or quadrupole motion leads to the well-known
Thouless-Valatin \cite{TV62} corrections to the mass, moment of
inertia, or to the vibrational mass parameters, respectively.

These corrections reflect the fact that a velocity-dependent
mean field should be appropriately transformed to the intrinsic
frame of reference \cite{BM75} corresponding to a given collective mode.
In fact, simple estimations and numerical calculations
\cite{Mig59,GQ80,DS81} show that the inertia obtained for
effective forces without the effective-mass terms ($m^*$=$m$)
are very close to those for $m^*$$<$$m$ when the
Thouless-Valatin corrections are consistently included.

Fast nuclear rotation is a phenomenon in which the collective
motion should be described beyond the adiabatic
limit. Such a cranking model has been
successfully used in explaining numerous high-spin effects in
nuclei \cite{Szy83}. In this approach, the properties of the
rotating mean field explicitly depend on the angular velocity.
However, most studies performed up to now were done in terms of
phenomenological mean fields, which are not selfconsistently
depending on the rotating states, and, therefore, do not
incorporate time-odd terms.  Only in Ref.{} \cite{NPF76} an
attempt has been made to include the effects of a zero-range
interaction within the Nilsson single-particle mean field.  Full
selfconsistent cranking calculations
\cite{BON87a,BON91a,CHB92,GBD94,Flo94,THB95,Hee95,Bon98,KR89,KR93,ER93}
are still rather scarce,
and no explicit analysis of the time-odd mean-field components
is available.

On the other hand, the discovery of the superdeformation
\cite{JK91}, and the resulting avalanche of very precise data on
high-spin states, allow for an attempt to study these unknown
aspects of the nuclear mean field. In particular, the phenomenon
of identical bands (see Ref.{} \cite{BHN95} for a recent
review) provides an extremely rich and puzzling information on
properties of fast rotating states. In terms of the mean field
approaches explicitly depending on the time-odd components,
the calculated identical $\gamma$-ray transitions
have been obtained in two cases only, namely, (i) the HF cranking
calculations with Skyrme interaction gave yrast band of
$^{194}$Hg identical to an excited band in $^{194}$Pb
\cite{CHB92}, and (ii) the identical bands in $^{152}$Dy and $^{151}$Tb
(excited band) were obtained in the relativistic mean-field
(RMF) theory \cite{KR93}.  In both studies, the authors invoked
the time-odd components of the mean field as a crucial element
of the obtained results.

In the present study we aim at a detailed analysis of these
time-odd terms in the context of the identical bands phenomenon.
We focus our attention on two classic and experimentally well
studied pairs of identical bands, namely, on those in $^{152}$Dy and
$^{151}$Tb (excited band), and in
$^{151}$Tb and $^{150}$Gd (excited band). We
perform our calculations in terms of the HF cranking method with
the Skyrme interactions.

It was not in our intentions to give here a full account of the
time-reversal breaking description that the selconsistent HF cranking
approach really offers and the experiment may test. Such a field
of studies seems to emerge now in relation with the fast progress in
the available experimental information. In particular, our choice
of the illustrative material purposely minimizes rather than
maximizes a possible magnitude of the time-odd effects. This is so,
because we mainly discuss the role of the $\pi[301]{1/2}$
Nilsson orbital known to interact only weakly via the Coriolis term
in the hamiltonian.

In all standard approaches, such as the Woods-Saxon (WS), Nilsson,
HF, or RMF cranking models,
one signature member of this orbital is to a good approximation
given by a staight line as a function of the angular velocity.
This type of dependence can be opposed to dramatic changes
of other aligning or interacting orbitals which are much more
sensitive to the rotating field and hence to its time-odd components.
On the other hand, the specific behavior of the discussed orbital
induces a regular behavior of various observables, and has a great
advantage in the fact that the level repulsion or level crosings
do not disturb the comparisons of primary interest here.

In Section~\ref{sec3} we discuss the Skyrme
interaction and energy density with a particular emphasis on the
relations between the time-even and time-odd components of the
mean field.  The HF cranking calculations are presented in
Section~\ref{sec4}, where several variants of the Skyrme functional
are used and the influence of the
time-odd terms on the rotational properties is
analyzed, and conclusions are presented in Section~\ref{sec6}.

\section{Skyrme energy density}
\label{sec3}

The Skyrme energy functional \cite{VB72,EBG75} is a
three-dimensional integral
   \begin{equation}\label{eq107}
   {\cal E} = \int d^3\bbox{r} {\cal H}(\bbox{r})
   \end{equation}
of the energy density ${\cal H}(\bbox{r})$ which can be
represented as a sum of the kinetic energy and of the
potential-energy isoscalar and isovector terms
   \begin{equation}\label{eq108}
   {\cal H}(\bbox{r}) = \frac{\hbar^2}{2m}\tau_0
               + {\cal H}_0(\bbox{r})
               + {\cal H}_1(\bbox{r}) ,
   \end{equation}
where
   \begin{equation}\label{eq109a}
   {\cal H}_t(\bbox{r})  =  {\cal H}^{\text{even}}_t(\bbox{r})
                         +  {\cal H}^{\text{odd}}_t (\bbox{r}) ,
   \end{equation}
with
\widetext
   \begin{equation}\label{eq109}\begin{array}{rclclclclcl}
   {\cal H}^{\text{even}}_t(\bbox{r})
    &=& C_t^{\rho}            \rho_t^2
    &+& C_t^{\Delta\rho}      \rho_t\Delta\rho_t
    &+& C_t^{\tau}            \rho_t\tau_t
    &+& C_t^{   J}            \tensor{J}_t^2
    &+& C_t^{\nabla J}        \rho_t\bbox{\nabla}\cdot\bbox{J}_t,
                                                      \rule{0ex}{4ex} \\
   {\cal H}^{\text{odd}}_t(\bbox{r})
    &=& C_t^{   s}            \bbox{s}_t^2
    &+& C_t^{\Delta s}        \bbox{s}_t\cdot\Delta
                              \bbox{s}_t
    &+& C_t^{   T}            \bbox{s}_t\cdot\bbox{T}_t
    &+& C_t^{   j}            \bbox{j}_t^2
    &+& C_t^{\nabla j}        \bbox{s}_t\cdot(\bbox{\nabla}\times\bbox{j}_t),
                                                       \rule{0ex}{4ex} \\
   \end{array}\end{equation}%
\narrowtext\noindent%
and the isospin index $t$ can have values 0 or 1.  Coupling
constants $C_t$ have superscripts corresponding to various
density functions appearing in Eq.~(\ref{eq109}).  All of these
coupling constants can, in principle, depend on particle densities,
see Refs.{} \cite{KKS77,PRL94}, but the most common choice
\cite{VB72,Bei75} restricts the density dependence to the
$C_t^{\rho}$ and $C_t^s$ terms.  In Appendix~\ref{appA} we give
coupling constants $C_t$ expressed through the traditional
parameters of the Skyrme interaction.

Apart from the density-dependent coupling constants, the
isoscalar energy density, ${\cal H}_0(\bbox{r})$, depends on the
isoscalar density functions and the isovector one, ${\cal
H}_1(\bbox{r})$, depends on the isovector density functions.
For the particle densities, $\rho_t$, the isoscalar and
isovector parts are defined in the usual way as a sum and a
difference of the proton and neutron contributions,
respectively,
   \begin{equation}\label{eq110}
   \rho_0 = \rho_n + \rho_p \quad , \quad
   \rho_1 = \rho_n - \rho_p ,
   \end{equation}
and analogous expressions are used to define other densities.

Altogether we have to consider six position-dependent density
functions $\rho$, $\tau$, $\bbox{j}$, $\bbox{s}$, $\bbox{T}$ and
$\tensor{J}$ with definitions given in Ref.{} \cite{EBG75}.
Whenever the isospin indices are omitted, we understand that the
symbols may refer either to the isoscalar ($t$=0) or isovector
($t$=1) parts.

In the energy densities (\ref{eq109}) there appear two scalar
time-even density functions, $\rho$ and $\tau$, one vector
time-odd, $\bbox{j}$, and two pseudo-vector time-odd ones,
$\bbox{s}$ and $\bbox{T}$, and one pseudo-tensor time-even
density function, $\tensor{J}$.  The vector time-even density
$\bbox{J}$ which appears in Eq.~(\ref{eq109}) is given by the
antisymmetric part of the pseudo-tensor density, i.e.,
$J_\lambda$=$\sum_{\mu\nu}\epsilon_{\lambda\mu\nu}J_{\mu\nu}$,
and is not an independent quantity.  Terms in ${\cal
H}^{\text{even}}_t(\bbox{r})$ and ${\cal
H}^{\text{odd}}_t(\bbox{r})$ are bilinear in time-even and
time-odd densities, respectively. Therefore, we denote them by
the superscripts ``even'' and ``odd''.  This notation is only
used to indicate the dependence on two different classes of
densities. It should not be confused with the fact that both
${\cal H}^{\text{even}}_t(\bbox{r})$ and ${\cal
H}^{\text{odd}}_t(\bbox{r})$ are of course {\em even} with
respect to the time reversal.  To every term in ${\cal
H}^{\text{even}}_t(\bbox{r})$ there corresponds an analogous
term in ${\cal H}^{\text{odd}}_t(\bbox{r})$, as seen in
Eq.~(\ref{eq109}).

\subsection{Mean fields}
\label{sec2a}

By varying the energy density (\ref{eq109}) with respect to the
six density functions $\rho$, $\tau$, $\bbox{j}$, $\bbox{s}$,
$\bbox{T}$ and $\tensor{J}$ one obtains the mean fields. Details
of calculations are presented in Ref.{} \cite{EBG75}. Here we
only repeat the final results conforming to the notation
introduced above.  Time-even and time-odd mean fields are
obtained by a variation of ${\cal H}^{\text{even}}_t(\bbox{r})$
and ${\cal H}^{\text{odd}}_t(\bbox{r})$, respectively, and they
read
\widetext
   \begin{equation}\label{eq209}\begin{array}{rclclcl}
   {\Gamma}^{\text{even}}_t
    &=& -\bbox{\nabla}\cdot M_t(\bbox{r})\bbox{\nabla}
    &+& U_t(\bbox{r})
    &+& \frac{1}{2i}\Big(\tensor{\nabla\sigma}\cdot\tensor{B}_t(\bbox{r})
                        + \tensor{B}_t(\bbox{r})\cdot\tensor{\nabla\sigma}
                    \Big),                               \rule{0ex}{4ex} \\
   {\Gamma}^{\text{odd}}_t
    &=& -\bbox{\nabla}\cdot\Big(\bbox{\sigma}\cdot\bbox{C}_t(\bbox{r})\Big)
         \bbox{\nabla}
    &+& \bbox{\sigma}\cdot\bbox{\Sigma}_t(\bbox{r})
    &+& \frac{1}{2i}\Big(\bbox{\nabla}\cdot\bbox{I}_t(\bbox{r})
                        + \bbox{I}_t(\bbox{r})\cdot\bbox{\nabla}\Big).
                                                          \rule{0ex}{4ex} \\
   \end{array}\end{equation}%
\narrowtext\noindent%
These fields are given by the six potential functions $U$, $M$,
$\bbox{I}$, $\bbox{\Sigma}$, $\bbox{C}$ and $\tensor{B}$ which
have tensor transformation properties respectively identical to
the six density functions on which they depend through the
following formulae
   \begin{mathletters}\begin{eqnarray}
   U_t
       &=& 2C_t^{\rho}           \rho_t
        +  2C_t^{\Delta\rho}      \Delta\rho_t
        +  C_t^{\tau}            \tau_t
        +  C_t^{\nabla J}        \bbox{\nabla}\cdot\bbox{J}_t,
                                                            \label{eq210a} \\
   \bbox{\Sigma}_t
       &=& 2C_t^{   s}           \bbox{s}_t
        +  2C_t^{\Delta s}        \Delta\bbox{s}_t
        +  C_t^{   T}            \bbox{T}_t
        +  C_t^{\nabla j}        \bbox{\nabla}\times\bbox{j}_t,
                                                            \label{eq210b} \\
   M_t
       &=& C_t^{\tau}            \rho_t,                    \label{eq210c} \\
   \bbox{C}_t
       &=& C_t^{   T}            \bbox{s}_t,                \label{eq210d} \\
   \tensor{B}_t
       &=& 2C_t^{   J}           \tensor{J}_t
        -  C_t^{\nabla J}        \tensor{\nabla}\rho_t,     \label{eq210e} \\
   \bbox{I}_t
       &=& 2C_t^{   j}           \bbox{j}_t
        +  C_t^{\nabla j}        \bbox{\nabla}\times\bbox{s}_t,
                                                            \label{eq210f}
   \end{eqnarray}\end{mathletters}%
The tensor gradient operators in Eqs.~(\ref{eq209}) and
(\ref{eq210e}) are defined \cite{EBG75} as
$(\tensor{\nabla\sigma})_{\mu\nu}$=$\nabla_\mu\sigma_\nu$ and
$\nabla_{\mu\nu}$$=$$\sum_{\lambda}\epsilon_{\mu\nu\lambda}\nabla_\lambda$.

Due to the density dependence of the coupling constants
$C_t^{\rho}$ and $C_t^{   s}$ one has to add to the isoscalar
potential energy $U_0$ the rearrangement terms \cite{RS80} resulting from
the variation of these coupling constants with respect to the
isoscalar particle density, i.e.,
    \begin{equation}\label{eq211}
    U'_0 = \sum_t\left(
           \frac{\partial C_t^{\rho}}{\partial\rho_0}\rho_t^2
         + \frac{\partial C_t^{   s}}{\partial\rho_0}\bbox{s}_t^2\right).
    \end{equation}
These terms introduce an explicit dependence of the complete
time-even isoscalar potential $U_0$+$U'_0$ on the isovector
density $\rho_1$ and on the time-odd densities $\bbox{s}_t^2$.
Similar terms have to be also consistently taken into account
whenever some other coupling constants are assumed to be density
dependent.

Finally, the neutron and proton hamiltonians, $h_n$ and $h_p$,
are obtained by combining the kinetic energy with the isoscalar
and isovector mean fields:
    \begin{equation}\begin{array}{rcl}
    h_n &=& -\frac{\hbar^2}{2m}\Delta
         +  {\Gamma}^{\text{even}}_0 + {\Gamma}^{\text{odd}}_0
         +  {\Gamma}^{\text{even}}_1 + {\Gamma}^{\text{odd}}_1,
                                                           \rule{0ex}{4.5ex} \\
    h_p &=& -\frac{\hbar^2}{2m}\Delta
         +  {\Gamma}^{\text{even}}_0 + {\Gamma}^{\text{odd}}_0
         -  {\Gamma}^{\text{even}}_1 - {\Gamma}^{\text{odd}}_1.
                                                           \rule{0ex}{4.5ex}
    \end{array}\end{equation}

We are now in a position to discuss different time-odd terms in
the mean fields. It is clear that the time-odd mean fields,
${\Gamma}^{\text{odd}}_t$, in
Eq.~(\ref{eq209}) directly result from the ``odd'' part,
${\cal H}^{\text{odd}}_t(\bbox{r})$, of the
energy density in (\ref{eq109}), and depend on 10 time-odd
coupling constants
$C_t^{   s}$,
$C_t^{\Delta s}$,
$C_t^{   T}$,
$C_t^{   j}$, and
$C_t^{\nabla j}$ for $t$=0 and $t$=1.
Similarly, the time-even mean fields depend on 10 time-even
coupling constants
$C_t^{\rho}$,
$C_t^{\Delta\rho}$,
$C_t^{\tau}$,
$C_t^{   J}$, and
$C_t^{\nabla J}$.
For the Skyrme interaction, the time-odd coupling constants are
linear combinations of the time-even ones (see Appendix~\ref{appA}),
and therefore the time-odd mean fields are uniquely determined
from the time-even mean fields.  Since the time-even fields are
tested against numerous experimental observations which have a
static character, they are much better known then the time-odd
ones. In fact, the Skyrme force parameters have been in the past
almost uniquely fitted to the static properties only.  A
description of dynamic properties which do depend on the
time-odd fields does not therefore (for the Skyrme force)
require new parameters to be introduced and fitted.

On the other hand, one sometimes adopts a different point of
view by considering the energy density to be a more fundamental
construction than the Skyrme interaction itself \cite{NV72}. In
such a case, all 20 coupling constants of Eq.~(\ref{eq109}) should
be treated and adjusted independently. However, in the next
Section we show that some relations between time-odd and time-even
coupling constants have an origin in the local gauge invariance
of the energy density.

\subsection{Local gauge invariance}
\label{sec2}

As noted in Ref.{} \cite{EBG75}, in the energy density
derived from the Skyrme interaction the kinetic density $\tau$
and the current density $\bbox{j}$ appear in the characteristic
combination of $\rho_t\tau_t$$-$$\bbox{j}_t^2$. The same is true
for two other pairs of densities appearing together in the
combinations
$\bbox{s}_t$$\cdot$$\bbox{T}_t$$-$$\tensor{J}_t^2$
and
$\Big(\rho_t\bbox{\nabla}$$\cdot$$\bbox{J}_t$$+$$\bbox{s}_t\cdot
                             (\bbox{\nabla}$$\times$$\bbox{j}_t)\Big)$.
This gives the following relations between three pairs of
time-even and time-odd coupling constants
   \begin{mathletters}\begin{eqnarray}
   C_t^{       j}  &=&  -C_t^{    \tau} ,\label{eq215a}  \\
   C_t^{       J}  &=&  -C_t^{       T} ,\label{eq215b} \\
   C_t^{\nabla j}  &=&  +C_t^{\nabla J} .\label{eq215c}
   \end{eqnarray}\end{mathletters}%
The Skyrme functional has now the form
   \begin{eqnarray}\label{eq109c}
   {\cal H}_t(\bbox{r})
    &=& C_t^{\rho}            \rho_t^2
     +  C_t^{   s}            \bbox{s}_t^2
     +  C_t^{\Delta\rho}      \rho_t\Delta\rho_t
     +  C_t^{\Delta s}        \bbox{s}_t\cdot\Delta\bbox{s}_t \nonumber \\
    &+& C_t^{\tau}      \left(\rho_t\tau_t-\bbox{j}_t^2  \right)
     +  C_t^{   T}      \Big(\bbox{s}_t\cdot
                              \bbox{T}_t - \tensor{J}_t^2\Big) \nonumber  \\
    &+& C_t^{\nabla J}  \Big(\rho_t\bbox{\nabla}\cdot\bbox{J}_t
                            + \bbox{s}_t\cdot
                             (\bbox{\nabla}\times\bbox{j}_t)\Big) .
   \end{eqnarray}

In Ref.{} \cite{EBG75} this structure has been interpreted as
a result of the Galilean invariance of the Skyrme interaction.
However, it has a deeper origin in the fact that the Skyrme
force is locally gauge invariant.  In order to illustrate the
role of the locally gauge invariant, velocity dependent
interactions let us consider an arbitrary finite-range and
non-local, but velocity independent interaction given by
   \begin{equation}\label{e201}
   \hat V = V(\bbox{r}'_1,\bbox{r}'_2;\bbox{r}_1,\bbox{r}_2) .
   \end{equation}
To simplify the notation we disregard for a moment the spin and
isospin variables. $\hat V$ describes an interaction process
where the particles 1 and 2 are located at $\bbox{r}_1$ and
$\bbox{r}_2$ before the interaction, and at $\bbox{r}'_1$ and
$\bbox{r}'_2$ after the interaction. When the system of
particles is described by a one-body density matrix
$\rho(\bbox{r},\bbox{r}')$, its HF interaction energy reads
   \begin{eqnarray}
   {\cal E}^{\text{int}}
    &=& \int d\bbox{r}'_1 d\bbox{r}'_2 d\bbox{r}_1 d\bbox{r}_2
            V(\bbox{r}'_1, \bbox{r}'_2; \bbox{r}_1, \bbox{r}_2) \times
                                                              \nonumber \\
     && \big(\rho(\bbox{r}_1,\bbox{r}'_1)\rho(\bbox{r}_2,\bbox{r}'_2)
      -      \rho(\bbox{r}_2,\bbox{r}'_1)\rho(\bbox{r}_1,\bbox{r}'_2)\big) .
                                                              \label{e202}
   \end{eqnarray}
{}For a local gauge transformation of the many-body HF wave
function $|\Psi\rangle$,
   \begin{equation}\label{eq111}
   |\Psi'\rangle = \exp\left\{i\sum_{j=1}^A\phi(\bbox{r}_j)\right\}
                    |\Psi\rangle,
   \end{equation}
where $\phi(\bbox{r})$ is an arbitrary real function of the
position $\bbox{r}$, one obtains the following gauge-transformed
one-body density matrix
   \begin{equation}\label{eq203}
   \rho'(\bbox{r},\bbox{r}')=
     \exp\left\{i\Big(\phi(\bbox{r})-\phi(\bbox{r}')\Big)\right\}
                                       \rho(\bbox{r},\bbox{r}') .
   \end{equation}

The general interaction energy (\ref{e202}) is not invariant
with respect to such a transformation. However, when the
interaction is local \cite{RS80},
   \FL\begin{equation}\label{e204}
   V(\bbox{r}'_1,\bbox{r}'_2;\bbox{r}_1,\bbox{r}_2) =
     \delta(\bbox{r}'_1\!-\!\bbox{r}_1)\delta(\bbox{r}'_2\!-\!\bbox{r}_2)
   V(\bbox{r}_1,\bbox{r}_2) ,
   \end{equation}
the corresponding interaction energy,
   \begin{eqnarray}
   {\cal E}^{\text{int}}
    &=& \int d\bbox{r}_1 d\bbox{r}_2
            V(\bbox{r}_1, \bbox{r}_2)
                                                   \times  \nonumber \\ &&
        \Big(\rho(\bbox{r}_1)\rho(\bbox{r}_2)
      -      \rho(\bbox{r}_2,\bbox{r}_1)\rho(\bbox{r}_1,\bbox{r}_2)\Big) ,
                                                              \label{e205}
   \end{eqnarray}
becomes invariant with respect to the local gauge.  The direct
term is invariant because it depends only on the gauge-invariant
local densities, which are denoted by the single argument, i.e.,
$\rho(\bbox{r})$$\equiv$$\rho(\bbox{r},\bbox{r})$. On the other
hand, in the exchange term the gauge factors coming from two
density matrices cancel one another.

When the density functions defining the Skyrme energy functional
(\ref{eq109}) are calculated for the gauge-transformed density
matrix (\ref{eq203}) one obtains the following relations:
   \begin{mathletters}\begin{eqnarray}
   \rho' &=& \rho
                                                   \label{eq112a}  \\
   \tau' &=& \tau + 2\bbox{j}\cdot
                     \bbox{\nabla}\phi
                  +  \rho(\bbox{\nabla}\phi)^2 ,
                                                   \label{eq112b}  \\
    s'_k &=& s_k ,
                                                   \label{eq112c}  \\
    j'_k &=& j_k  +  \rho\nabla_k\phi ,
                                                   \label{eq112d}  \\
    T'_k &=& T_k  + 2\sum_lJ_{kl}\nabla_l\phi ,
                  +  s_k(\bbox{\nabla}\phi)^2 ,
                                                   \label{eq112e}  \\
  J'_{kl}&=& J_{kl}  +  s_l\nabla_k\phi ,
                                                   \label{eq112f}
   \end{eqnarray}\end{mathletters}%
and the three characteristic combinations of density functions,
which appear in the energy density of Eq.~(\ref{eq109c}), are then
explicitly gauge invariant.
Transformation properties of $\tau$ and $\bbox{j}$ allow to
interpret $\bbox{\nabla}\phi$ as a velocity field,
   \begin{equation}\label{eq212}
   \bbox{v} = \frac{\hbar}{m}\bbox{\nabla}\phi,
   \end{equation}
which shows that the
flow of matter obtained through the gauge transformation is
irrotational, $\bbox{\nabla}$$\times$$\bbox{v}$=0.

The local gauge invariance of the Skyrme interaction reflects
the fact that its velocity dependence has been introduced only
to simulate the finite range effects of the effective
interaction.  In this way the Skyrme interaction conserves the
local gauge invariance of a velocity-independent finite-range
local interaction, such as the Gogny force \cite{RS80}, for example.

\subsubsection{Translational motion}\label{sec2aa}
The Galilean invariance is a special case of the local gauge
invariance, for which the phase in Eq.~(\ref{eq111}) is given by
   \begin{equation}\label{eq113}
   \phi(\bbox{r}) = \frac{\bbox{p}\cdot\bbox{r}}{\hbar} ,
   \end{equation}
where $\bbox{p}$ is a constant linear momentum of the boost
transformation. Since the interaction energy does not depend on
$\bbox{p}$, the total energy (\ref{eq107}) is only modified
through the kinetic energy [the first term in Eq.~(\ref{eq108})].
Using the transformation property of $\tau$, Eq.~(\ref{eq112b}), we
have the energy increase under the boost transformation,
   \begin{equation}\label{eq206}
   \Delta{\cal E}^{\text{boost}} = \frac{\bbox{p}^2}{2m}A ,
   \end{equation}
equal to the translational energy of the boosted system. This
result holds for an initially stationary solution
(i.e., for vanishing currents, $\bbox{j}$=0),
however, due to the transformation property of
$\bbox{j}$, Eq.~(\ref{eq112d}), the boost transformations can be
added to one another by adding the corresponding momenta
$\bbox{p}$.
{}For the boost transformation, one obtains the obvious
velocity field (\ref{eq212}), i.e.,
   \begin{equation}\label{eq213a}
   \bbox{v}^{\text{boost}} = \frac{\bbox{p}}{m} .
   \end{equation}

\subsubsection{Rotational motion}\label{sec2ab}
Since the velocity field (\ref{eq212}) obtained through a
gauge transformation is
irrotational, it cannot correctly describe physical rotations of nuclei.
This is so, because the nuclei basically rotate as rigid bodies
(at least in the independent-particle approximation \cite{BM75}),
and the velocity field of a rigid-body rotation,
$\bbox{v}^{\text{rigid}}$=$\bbox{\omega}\times\bbox{r}$, has a non-zero
curl, $\bbox{\nabla}$$\times$$\bbox{v}^{\text{rigid}}$=2$\bbox{\omega}$,
i.e., is not irrotational. Of course, interactions and
nucleon-nucleon correlations
(pairing) introduce an irrotational component in the velocity field
(moment of inertia decreases below the rigid-body value), but this field
is never entirely irrotational.

In an analogy to the boost transformation, one may try to
induce the rotation of a many-fermion system by adding for all
particles a constant value, $j_x$,
to their projections of the angular momentum on a fixed (say $x$) axis.
Such a procedure can be realized in terms of the twirl
transformation given by the gauge function
   \begin{equation}\label{eq114}
   \phi(\bbox{r}) = \frac{j_x{\arctan}(z/y)}{\hbar} .
   \end{equation}
Its velocity field has the form
   \begin{equation}\label{eq213}
   \bbox{v}^{\text{twirl}} = \frac{j_x}{m\omega\eta(y,z)^2}
                             \bbox{\omega}\times\bbox{r},
   \end{equation}
where $\bbox{\omega}$ is the vector of angular velocity oriented along
the $x$ axis, and $\eta(y,z)$ is the distance to this axis.  One can
see that the velocity field (\ref{eq213}) is singular at the
rotation axis. The energy increase resulting from the kinetic
energy density (\ref{eq112b}) is then infinite for any
arbitrarily small $j_x$. The irrotational velocity field
(\ref{eq213}) is of course very much different from the
rigid-rotation velocity field, $\bbox{v}^{\text{rigid}}$, even though
both contain the same factor $\bbox{\omega}\times\bbox{r}$.

This illustrative example shows that the nuclear rotation cannot
be realized by democratically distributing the angular momentum
among all particles. In the cranking approximation, at a given
value of the angular velocity some particles
receive larger contributions (aligning states) and some smaller
contributions (high-$K$ states). A precise distribution cannot
be found without actually solving the quantal cranking equations.
This example also shows that the gauge-invariant interaction
must contribute to the rotational energy, contrary to what
happens in the case of the translational motion.

\section{Hartree-Fock cranking calculations}
\label{sec4}

In the present study we have performed the Hartree-Fock (HF)
calculations of superdeformed rotational bands using the Skyrme
effective interactions.  The calculations have been done using
the numerical code HFODD which employs a three-dimensional
cartesian deformed harmonic oscillator (HO) basis to describe
the single-particle wave functions.  The details concerning the
HFODD code will be presented elsewhere \cite{J2D2a}; here we
only give a few of its basic parameters pertaining to the
present application.

The calculations have been performed using a fixed basis given
by the HO frequencies ${\hbar\omega}_{\perp}$=11.200 and
${\hbar\omega}_{\parallel}$=6.246\,MeV in the directions perpendicular and
parallel to the symmetry axis, respectively. These values have
been obtained by standard prescriptions developed for
diagonalizing the deformed WS hamiltonian
\cite{DST81,CDN87} in the HO basis, and correspond to the WS
potential with deformations $\beta_2$=0.61 and $\beta_4$=0.10.
The basis has been restricted to a fixed number, $M$, of basis states
having the lowest single-particle HO energies
$\epsilon_{HO}$=$(n_x$+$n_y$+$1){\hbar\omega}_{\perp}
$+$(n_z$+$\case{1}{2}){\hbar\omega}_{\parallel}$.  The actual calculations
have been performed with $M$=306. This corresponds to the
maximal numbers of oscillator quanta equal to 8 and 15 in the
perpendicular and parallel directions, respectively.

The stability of results with respect to increasing the size of
the HO basis has been tested by performing calculations with
$M$=604, which introduces basis states up to 11 and 20 quanta in
these two directions.  It has been found that the rotational
characteristics of the studied nuclei are almost independent of
such an increase. For example, numerical inaccuracies in the
dynamical moment ${{\cal J}^{(2)}}$ and in the total angular
momentum $I$ can be estimated to be smaller than
0.2\,{$\hbar^2$MeV$^{-1}$} and
0.1\,$\hbar$, respectively. Inaccuracies of relative values
between different angular frequencies $\omega$, or between
different nuclei, are smaller than these estimates, because the
numerical errors then cancel out.

In the present study we aim at investigating the role of
different time-odd terms in the selfconsistent mean fields
(\ref{eq209}) obtained for rotating superdeformed nuclei. As
discussed in Section~\ref{sec3}, this can be done by considering
different values of 10 time-odd coupling constants appearing in
the Skyrme energy density, Eq.~(\ref{eq109}).  For every given set
of values of coupling constants we perform full selfconsistent
calculations within the HF cranking method.  In the present
study, pairing correlations are not taken into account.  Below
we separately discuss three cases corresponding to (i) the
complete Skyrme functional, (ii) the Skyrme functional with
certain time-odd terms omitted, but with the gauge invariance
preserved, and (iii) with the gauge invariance violated.

As discussed in Section~\ref{sec2}, the energy functionals
corresponding to the standard Skyrme interactions preserve the
local gauge invariance.  The 20 coupling constants appearing in
Eq.~(\ref{eq109}) are then restricted by 6 conditions
(\ref{eq215a})-(\ref{eq215c}), which leads to the energy density
of Eq.~(\ref{eq109c}).  In Table~\ref{tab1}, the remaining 14 coupling
constants are listed for the SIII \cite{Bei75}, SkM*
\cite{BQB82}, and SkP \cite{DFT84} Skyrme interactions.  The
values of the density-dependent coupling constants $C_t^{\rho}$
and $C_t^{s}$ are given for the vacuum ($\rho$=0) and for the
nuclear-matter saturation density ($\rho$=$\rho_{\text{nm}}$)
characterizing a given force.

The time-even coupling constants corresponding to the SIII,
SkM*, and SkP interactions are rather similar. The main
difference consists in different values of the
isoscalar-effective-mass coupling constant $C_0^\tau$ which is
equal to 0 for SkP (effective mass $m^*/m$=1), and 44.4\,{MeV$\cdot$fm$^5$}
and 34.7\,{MeV$\cdot$fm$^5$} for SIII and SkM*, respectively (effective
masses $m^*/m$=0.76 and 0.79). Apart from that, the absolute
values of the $C_t^\rho$ coupling constants are larger for SkP
than for SIII and SkM*, which gives better symmetry-energy
properties \cite{Dob94a} within the SkP parametrization as
compared to the other two forces.  On the other hand, the SkM*
parameters have been adjusted so as to properly describe the
surface energy at large deformation, and therefore this
interaction was successfully applied in numerous studies of
superdeformation.

Values of the time-odd coupling constants corresponding to the
three Skyrme interactions, Table~\ref{tab1}, differ much more than
those of the time-even ones. This illustrates uncertainties in
determining the time-odd components of the mean field.  The
differences partly result from the fact that SkP has been
adjusted to give attractive matrix elements in the pairing
channel, while those given by SIII and SkM* are repulsive.  One
should bear in mind, however, that for the Skyrme interaction the
time-odd coupling constants are unique functions of the
time-even ones (see Appendix~\ref{appA}) and, in principle, one has
no freedom for an independent readjustment. Such a readjustment
is possible only if we consider the HF theory based directly on
the energy density functional and not on the Skyrme interaction.

In the previous applications of the Skyrme force to nuclear
rotation \cite{BON87a,BON91a,CHB92,GBD94}, in the energy density
(\ref{eq109c}) the terms $C_t^{\Delta s}$ and $C_t^T$ have been
neglected in order to facilitate the calculations. The first of
these terms gives purely time-odd contribution to the mean
field, while the second one gives both time-even and time-odd
contributions, because the gauge invariance implies that
$C_t^T$=$C_t^J$, Eq.~(\ref{eq215b}). In fact,
the term $C_t^J$ has also been usually neglected in most Skyrme
parametrizations applied to problems where the time-reversal symmetry
is concerved,
cf.{} Ref.{} \cite{Bei75}.  Since the HFODD code is
organized in a different way, omitting some terms would not
provide any serious simplifications, and the code may in fact
handle the complete Skyrme functional. This gives us a
possibility to test the importance of different terms for the
rotational properties of nuclei.

There exist several high-spin observables which, when calculated
from the HF solutions (wave functions) behave differently
depending on whether various time-odd terms are included or not.
This offers, in principle, a possibility of both better
readjustment of the interaction coupling constants
and better understanding of the underlying mechanisms. Some of the
physical quantities, as e.g.{} alignments and
${{\cal J}^{(2)}}$-moments of intruder
orbitals, are
well recognized as responding strongly to the Coriolis and centrifugal
interaction effects. The same quantities are expected to respond relatively
strongly to the time-odd terms in the mean-field hamiltonian.
Similarly, various families of orbitals having large high-$j$ components are
systematically responsible for such precisely measurable effects
as band-crossings (back- or up-bending effects) and related phenomena.

In this study we would like to illustrate the effects of the time-odd terms
on yet another seemingly more subtle mechanism related to identical
bands, leaving the aformentioned analysis of intruder orbitals
for a later investigation.

In the following Sections we present results of the HF cranking
calculations for the yrast superdeformed bands in dysprosium,
$^{152}$Dy(1), and terbium, $^{151}$Tb(1), and for the first excited bands
in the corresponding isotones, $^{151}$Tb(2) and $^{150}$Gd(2). According
to the standard notation, the numbers in parentheses refer to
numbers attributed in experimental studies in connection with
relative intensities of gamma transitions. The experimental data
are taken from Refs.{} \cite{Bea93,Dag94,Kha95} where the
most recent and precise results are given.

The pairs of superdeformed bands, $^{151}$Tb(2)--$^{152}$Dy(1) and
$^{150}$Gd(2)--$^{151}$Tb(1), are identical with a very high precision
\cite{Byr90}.  For each pair, the $\gamma$-ray transition
energies $E_\gamma$ are identical up to 2\,keV. The identity
of the bands can be characterized in two ways, (i) by their
relative alignments, and (ii) by their relative
dynamical moments.  The relative alignment $\delta{I}$ is defined as a
difference of spins in two bands at fixed angular velocity
$\omega$. Similarly, the relative dynamical moment
$\delta{{\cal J}^{(2)}}$ is the difference of
${{\cal J}^{(2)}}$ at the same value of
$\omega$.  In calculations, the latter is a derivative of the
former with respect to the angular velocity.

In the present study we have fixed the yrast proton
configurations of $^{152}$Dy and $^{151}$Tb to (16,16,17,17) and
(15,16,17,17), respectively.  The numbers in parentheses denote
the numbers of lowest states occupied in the parity-signature blocks
ordered as (++,+$-$,$-$+,$-$$-$), where the signs denote the
parity quantum number and the sign of the (imaginary) signature
\cite{BM75} quantum number, $r$=${\pm}i$.
One somtimes uses a notation based on the signature index $\alpha$
which equals to $-$1/2(+1/2) for $r$=+$i$($-i$). For all bands
considered in the present study, the neutron configurations are
fixed at (22,22,21,21).  The yrast configuration in $^{151}$Tb
corresponds to a hole in the 16-th orbital in the ++ block,
16$^{++}$, which in the standard notation is described as the
$\pi[651]{3/2}$~($r$=$+i$) or $\pi6_4$ Nilsson intruder orbital.

Since in the experiment the angular velocity is associated with
half of the transition energy $E_\gamma$, and the above pairs of
bands have identical transition energies, the relative
alignments must be close to a half-integer value.  This is
simply a consequence of the fact that the spins in the odd and
even nuclei are half-integer and integer, respectively.
Departures from half-integer relative alignments can only be
caused by differences in the $\gamma$-ray transition energies,
which are very small for the two pairs of bands studied here.
Since the values of spins have not yet been measured, the
relative alignments are known up to an additive integer value
and a theoretical input is necessary if one wishes to put
forward one value or another.

Already at a very early stage of the identical band studies,
it has been
suggested \cite{Naz90} that the first excited bands $^{151}$Tb(2) and
$^{150}$Gd(2) correspond to the
$\pi[301]{1/2}$~($r$=$+i$) holes (signature index
$\alpha$=$-$1/2) in the yrast states of the respective cores
(see also Ref.{} \cite{Rag93}).  In the present study we have
followed this interpretation and we have constructed the excited
bands by creating a hole in the 17-th orbital of the $-$+ block,
17$^{-+}$.

In all the WS and HF cranking calculations, the single-particle
routhian $\pi[301]{1/2}$~($r$=$+i$) increases with rotational
frequency with a
constant slope of about +0.5\,$\hbar$, i.e., it has the
single-particle alignment (the average value of the projection
of angular momentum on the rotation axis) close to
$-$0.5\,$\hbar$.  Therefore, a hole created in this orbital must
lead to relative alignment of about +0.5\,$\hbar$. In this paper
we adopt this half-integer value for fixing the unknown integer
additive constant, $\delta{I}_0$=+0.5\,$\hbar$, required to
extract relative alignment from experimental data. Within this
choice, the experimental average relative alignment for the
$^{151}$Tb(2)--$^{152}$Dy(1) pair of bands equals to +0.564(18)\,$\hbar$.
The error given here is the average error resulting from
experimental errors of transition energies.  The experimental
relative alignment for the $^{150}$Gd(2)--$^{151}$Tb(1) pair increases
slightly with the angular velocity and has the average value of
+0.479(14)\,$\hbar$.  In fact, experimentally, only the average
departures from the half-integer constant $\delta{I}_0$,
$\langle\delta{I}\rangle$=$\delta{I}_0${+}0.064(18)\,$\hbar$ and
$\langle\delta{I}\rangle$=$\delta{I}_0$$-$0.021(14)\,$\hbar$,
are established.

\subsection{Complete Skyrme functionals}
\label{sec4a}

In Figs.~\ref{fig05} and \ref{fig06} we show results of
calculations for the $^{151}$Tb(2)--$^{152}$Dy(1) and
$^{150}$Gd(2)--$^{151}$Tb(1)
pairs of bands, respectively.  The complete Skyrme
energy-density functionals of the SIII, SkM*, and SkP forces
have been used.  Parts (a) of these Figures (bottom) present the
dynamical moments ${{\cal J}^{(2)}}$ as functions of the angular
velocity $\omega$.

\subsubsection{Dynamical moments ${{\cal J}^{(2)}}$}
\label{sec4aa}

{}For $^{152}$Dy(1), Fig.~\ref{fig05}, all forces overestimate the
experimental values of ${{\cal J}^{(2)}}$ by
about 5-10\%. All three forces give very similar results, within
2\,{$\hbar^2$MeV$^{-1}$}.
A similarity of results obtained for different forces is also
visible in Table~\ref{tab6}, where we give the values of proton
quadrupole moments calculated at $\omega$=0.5\,{MeV/$\hbar$}.  The
SIII interaction gives values larger by only 0.3--0.4\,b as
compared to those given by SkM*, while the SkP force leads to
intermediate results. A similarity of results for changes of
proton quadrupole moments induced by creating holes in proton
orbitals is even more pronounced. A polarization by the
$\pi[301]{1/2}$~($r$=$+i$) hole gives, for different forces,
the changes between
0.12\,b and 0.15\,b.  That induced by the
$\pi[651]{3/2}$~($r$=$+i$) hole gives
values between $-$0.94\,b and $-$1.05\,b.  The HF proton
quadrupole moment of $^{152}$Dy(1) agrees very well with the result
obtained within the Nilsson single-particle potential
\cite{Rag93}.

In Fig.~\ref{fig06}(a) we show the dynamical moments
${{\cal J}^{(2)}}$ calculated for the
$^{151}$Tb(1) band. Due to the hole in the
$\pi[651]{3/2}$~($r$=$+i$) intruder orbital,
${{\cal J}^{(2)}}$ decreases here with $\omega$
much faster than that of the $^{152}$Dy(1) band. However, at
$\omega$$\simeq$0.45\,{MeV/$\hbar$} one obtains values of
${{\cal J}^{(2)}}$$\simeq$94\,{$\hbar^2$MeV$^{-1}$} which are
almost identical for all the
forces and for both bands.  This contradicts simple perturbative
estimates. Indeed, a hole in the $^{152}$Dy(1) core should in
principle cause a decrease of the moment of inertia due to the smaller mass
and deformation of $^{151}$Tb(1), and also due to the fact that the
$\pi[651]{3/2}$~($r$=$+i$) routhian has negative second derivative
with respect
to $\omega$ and, therefore, the hole in this orbital should
bring negative perturbative contribution to ${{\cal J}^{(2)}}$.
Nevertheless, at $\omega$$\simeq$0.45\,{MeV/$\hbar$} the
${{\cal J}^{(2)}}$ values
calculated for $^{152}$Dy(1) and $^{151}$Tb(1) are the same. This
illustrates the fact that the polarization effects obtained by
selfconsistent calculations do not necessarily follow
perturbative estimates.

In $^{151}$Tb(1), the SIII and SkM* forces give a better agreement
with data than SkP.  This can be attributed to a different
effective masses, $m^*/m$=1 for SkP and 0.76--0.79 for SIII and
SkM*, which leads to different time-odd components in the mean
field (see the next Section). Then, the interaction
between the intruder orbital $\pi[651]{3/2}$~($r$=$+i$)
and the time-odd mean field is
modified, and gives a more significant departure from
experiment.

\subsubsection{Relative dynamical moments $\delta{{\cal J}^{(2)}}$}
\label{sec4ab}

In parts (b) of Figs.~\ref{fig05} and \ref{fig06} we present the
relative dynamical moments $\delta{{\cal J}^{(2)}}$ calculated for
the pairs of bands $^{151}$Tb(2)--$^{152}$Dy(1) and
$^{150}$Gd(2)--$^{151}$Tb(1),
respectively.  One should note that the scale in (b) is enlarged
five times with respect to that in (a). At
$\omega$$>$0.3\,{MeV/$\hbar$}, both pairs of bands have
dynamical moments identical up to about 1\,{$\hbar^2$MeV$^{-1}$},
with lighter
isotones having slightly larger values.  In the scale of (a)
this would lead to curves identical up to the size of the data
marks.

This result confirms the observation \cite{CHB92} that the
identity of dynamical moments can in fact be obtained in
self-consistent theories, and does not necessarily follow semiclassical
\cite{BQB94} or perturbative estimates. The experimental values
of $\delta{{\cal J}^{(2)}}$ are not shown in parts (b), because they would
be scattered between
$\pm$1.5\,{$\hbar^2$MeV$^{-1}$} with errors between 1 and
2.5\,{$\hbar^2$MeV$^{-1}$}, i.e., in the scale of (b) they would cover the
whole presented region of $\delta{{\cal J}^{(2)}}$.

\subsubsection{Relative alignments $\delta{I}$}
\label{sec4ac}

The calculated relative alignments $\delta{I}$, shown in parts
(c) of Figs.~\ref{fig05} and \ref{fig06}, do not reproduce
experimental results with a sufficient precision. For the
$^{151}$Tb(2)--$^{152}$Dy(1) pair [Fig.~\ref{fig05}(c)], in the region of
$\omega$ where the data are available, one obtains a gradual
increase of $\delta{I}$ by almost 0.5\,$\hbar$. For the SIII and
SkP forces the value obtained at $\omega$=0.3\,{MeV/$\hbar$} is
correct, but for SkM* the whole curve is additionally shifted up
by about 0.5\,$\hbar$.  For the $^{150}$Gd(2)--$^{151}$Tb(1) calculations
this increase becomes smaller (SIII and SkP) and similar to the
small increase seen in the experimental data. However, the
values of $\delta{I}$ are still slightly (SIII and SkP) or
significantly (SkM*) too large.

Values of relative alignments can be translated into the
differences $\delta E_\gamma$ of $\gamma$-transition energies
between the two bands. Using a linear local expansion of spin as
function of the angular frequency we obtain that
   \begin{equation}\label{eq216}
    \delta E_\gamma\simeq 2\hbar\delta\omega
                   \simeq 2\hbar\frac{\delta{I}-\delta{I}_0}
                                     {{{\cal J}^{(2)}}} ,
   \end{equation}
where $\delta\omega$ is the difference of frequencies at spins
of physical states in two nuclei.  Hence the departures of
calculated relative alignments $\delta{I}$ from
$\delta{I}_0$=0.5\,$\hbar$ correspond to the values of $\delta
E_\gamma$ between 0 and 10\,keV, while the measured values are
between 0 and 2\,keV.  In the following two sections we present
calculations obtained with modified Skyrme functionals. In this
way we try to analyze the influence of the time-odd terms on the
relative alignments discussed here.

\subsection{Modified gauge-invariant Skyrme functionals}
\label{sec4b}

As discussed above, the gauge-invariance conditions
(\ref{eq215a})-(\ref{eq215c}) restrict values of 6 time-odd
coupling constants and leave a freedom to modify the values of
$C_t^{   s}$ and  $C_t^{\Delta s}$.  If one decides to depart
from the complete Skyrme-interaction functional, in which the
time-odd coupling constants are uniquely defined by the
time-even ones (Appendix~\ref{appA}), one may, in principle, use
arbitrary values of $C_t^{   s}$ and  $C_t^{\Delta s}$.
However, independent variations of these coupling constants have
never been considered in the literature, and their effects are,
up to now, unknown.
Therefore, in the present study we restrict our analysis to the
functionals in which one or more of the SkM* coupling constants,
Table~\ref{tab1}, are assumed to be equal to zero. Moreover, in
order to further restrict the number of possible variants, we
only consider simultaneous modifications of the isoscalar and
isovector coupling constants of a given species. In the frame of
gauge-invariant functionals, this leaves us a possibility of
putting $C_t^{   s}$=0 and/or $C_t^{\Delta s}$=0.

Another modified gauge-invariant functional can be obtained by
removing the term
$\bbox{s}_t$$\cdot$$\bbox{T}_t$$-$$\tensor{J}_t^2$,
i.e., by putting $C_t^{T}$=$C_t^{J}$=0, in accordance with
Eq.~(\ref{eq215b}).  This leads to a modification of time-odd {\em
and} time-even mean fields.  However, the term $\tensor{J}_t^2$
was anyhow neglected in most parametrizations o the Skyrme
forces used for time-even studies, and in particular in SkM*.
Therefore, below we discuss four possibilities corresponding to
(i) the complete functional, (ii) $C_t^{   T}$=$C_t^{   J}$=0,
(iii) $C_t^{\Delta s}$=$C_t^{T}$=$C_t^{   J}$=0, and then (iv)
$C_t^{ s}$=$C_t^{\Delta s}$=$C_t^{   T}$=$C_t^{   J}$=0.  The
results are presented in Figs.~\ref{fig07} and \ref{fig08} for
the pairs of bands $^{151}$Tb(2)--$^{152}$Dy(1) and
$^{150}$Gd(2)--$^{151}$Tb(1),
respectively.  We have studied several other possibilities, but
principal conclusions can be drawn from these four cases.

{}For $^{152}$Dy(1), the dynamical moments ${{\cal J}^{(2)}}$,
Fig.~\ref{fig07}(a), are (at high spin) sensitive only to the
$C_t^{   T}$=$C_t^{   J}$ coupling constants.  The influence of
$C_t^{\Delta s}$ and $C_t^{   s}$ is visible only at low spins,
and moreover, their effects have opposite signs and partially
cancel one another.  Removing from the functional the term
$\bbox{s}_t$$\cdot$$\bbox{T}_t$$-$$\tensor{J}_t^2$
decreases the dynamical moments, and, at the
high-spin-end of the band, brings the calculated value down to
the experimental result.  At lower spins one obtains a smaller
than previously overestimation of the data (by about 5\%).

A very interesting results is obtained for the relative
dynamical moments $\delta{{\cal J}^{(2)}}$, Fig.~\ref{fig07}(b).
For any studied
combination of the time-odd mean fields taken into account, the
dynamical moments in both nuclei are almost identical.
Even if the {\em values} of ${{\cal J}^{(2)}}$ depend on whether the term
$\bbox{s}_t$$\cdot$$\bbox{T}_t$$-$$\tensor{J}_t^2$
is taken into account or not, the {\em differences} of ${{\cal J}^{(2)}}$ do
not depend on it at all.  This is a rather general observation,
valid also for other cases of modified functionals discussed
below. It means that the identity of dynamical moments
is fairly independent of at least some of
the details of effective interactions,
and can probably be attributed to rather simple geometric
effects.

The relative alignments $\delta{I}$, Fig.~\ref{fig07}(c), very
strongly depend on the presence of the
$\bbox{s}_t$$\cdot$$\bbox{T}_t$$-$$\tensor{J}_t^2$
term and are almost independent of the $\bbox{s}_t^2$ and
$\bbox{s}_t$$\cdot$$\Delta\bbox{s}_t$ terms.  As soon as
$C_t^{T}$=$C_t^{   J}$=0, the value of $\delta{I}$ at
$\omega$=0.3\,{MeV/$\hbar$} comes down to the experimental result.  On
the other hand, the increase of $\delta{I}$ as function of the
angular frequency is still too large, and does not give the
experimental identity of the $\gamma$-transition energies,
Eq.~(\ref{eq216}).

In Fig.~\ref{fig08}, for the pairs of bands $^{150}$Gd(2)--$^{151}$Tb(1) we
present a similar analysis of the role of different time-odd
terms in the Skyrme functional. In this case, the influence of
the
$\bbox{s}_t$$\cdot$$\bbox{T}_t$$-$$\tensor{J}_t^2$
term on ${{\cal J}^{(2)}}$ is weaker than for the
$^{151}$Tb(2)--$^{152}$Dy(1) pair,
while its removal leads to a good agreement with experiment,
with only a slightly too slow decrease of ${{\cal J}^{(2)}}$ as function of
the angular velocity. At the same time the relative alignments
become much closer to experimental data, but an overestimation
by a few tenth of $\hbar$ persists at all angular frequencies.
The influence of the other two time-odd terms considered here,
$\bbox{s}_t^2$ and $\bbox{s}_t$$\cdot$$\Delta\bbox{s}_t$, is
much weaker, although a non-negligible influence on relative
alignments can be seen.

We conclude this section by noting that the importance of the
gauge-invariant time-odd terms for the rotational properties is
{\em not} simply related to the order of the given term. In fact, the
zero-order term $\bbox{s}_t^2$, which depends on the density of
spin, has a very small influence on the results. At the same
time, two second-order terms,
$\bbox{s}_t$$\cdot$$\Delta\bbox{s}_t$ and
$\bbox{s}_t$$\cdot$$\bbox{T}_t$$-$$\tensor{J}_t^2$,
which in the previous studies have been simultaneously
neglected, have a small and rather large influence,
respectively. On the other hand, the latter term, which in
principle cannot be neglected because of its magnitude, leads to
larger deviations from experimental data, as compared with
calculations which do disregard it.  Of course, the present
analysis is not sufficient for a more accurate derivation of the
magnitude of coupling constants from experiment.  This can only
be done by a simultaneous consideration of many different
available data, and moreover, should also involve a careful
readjustment of properties of the time-even components of the
mean fields.

\subsection{Modified gauge-violating Skyrme functionals}
\label{sec4c}

In this Section we present results of calculations for Skyrme
functionals with coupling constants which do not obey the
gauge-invariance conditions (\ref{eq215a})-(\ref{eq215c}).  In
order to simplify the discussion, the gauge-invariant time-odd
terms $\bbox{s}_t^2$ and $\bbox{s}_t$$\cdot$$\Delta\bbox{s}_t$
are disregarded, i.e., $C_t^{   s}$=$C_t^{\Delta s}$=0.
Similarly as in the previous Section, we only consider
simultaneous modifications of the isoscalar and isovector
coupling constants. This leaves us with a possibility of
modifying the time-odd coupling constants, $C_t^{   j}$,
$C_t^{T}$, and $C_t^{\nabla j}$, while leaving unchanged their
time-even gauge-invariance partners, $C_t^{\tau}$, $C_t^{   J}$,
and $C_t^{\nabla J}$.

{}First we have checked the separate role of the time-even and
time-odd parts in the term
$\bbox{s}_t$$\cdot$$\bbox{T}_t$$-$$\tensor{J}_t^2$
which has been discussed in the previous Section. It turned out
that the time-even part, $\tensor{J}_t^2$, influences the
results for rotational properties in a negligible way.
Therefore, for all practical purposes the results discussed
previously should be attributed to the time-odd part,
$\bbox{s}_t$$\cdot$$\bbox{T}_t$.  This allows us to concentrate
here on the two other time-odd terms, $\bbox{j}_t^2$ and
$\bbox{s}_t$$\cdot$$(\bbox{\nabla}$$\times$$\bbox{j}_t)$.

It is worth recalling at this point that in the gauge-invariant
energy functional (\ref{eq109c}) the $\bbox{j}_t^2$ term is
associated with the effective-mass term $\rho_t\tau_t$, and the
term $\bbox{s}_t$$\cdot$$(\bbox{\nabla}$$\times$$\bbox{j}_t)$
comes together with the spin-orbit term
$\rho_t$$(\bbox{\nabla}$$\cdot$$\bbox{J}_t)$.  Therefore,
the first one is absent in any theory which has the effective
mass $m^*$ equal to the free mass $m$.  This is the case for all
studies based on phenomenological single-particle potentials,
like the Nilsson or WS ones.

On the other hand, the spin-orbit term is crucially important
for a correct ordering of single-particle shells and is always
taken into account in realistic calculations.  However, the
studies based on phenomenological potentials always disregard
its time-odd gauge partner, and therefore should be considered as
gauge-violating approaches, similar to the selfconsistent
calculations considered in this Section.

In Figs.~\ref{fig09} and \ref{fig10} we present the results
obtained for the same pairs of superdeformed bands as considered
above, namely, for $^{151}$Tb(2)--$^{152}$Dy(1) and
$^{150}$Gd(2)--$^{151}$Tb(1),
respectively.  Open symbols refer to the functionals with
$C_t^{j}$=0 and squares to those with $C_t^{\nabla j}$=0.
Consequently, the full circles repeat here the same results as
those shown by full circles in Figs.~\ref{fig07} and \ref{fig08}.
On the other hand, open squares correspond to the Skyrme
functionals with {\em all} time-odd terms neglected.

As expected from the semiclassical analysis \cite{BQB94},
omitting the $\bbox{j}_t^2$ term leads to much smaller values of
${{\cal J}^{(2)}}$. The obtained decrease is of the order of 10-13\%,
depending on the value of the angular frequency, i.e., it is
much smaller than the decrease of about 40\% obtained from
non-selfconsistent estimates \cite{BQB94}.  With a very high
precision, this decrease is identical for both nuclei in pairs
differing by a hole in the $\pi[301]{1/2}$~($r$=$+i$) orbital.
Indeed, even if
the values of ${{\cal J}^{(2)}}$ change by as much as
13\,{$\hbar^2$MeV$^{-1}$}, the
relative dynamical moments, Figs.~\ref{fig09}(b) and
\ref{fig10}(b), do not change at all, and are always well below
1\,{$\hbar^2$MeV$^{-1}$}. In addition, the $\bbox{j}_t^2$ term has only a minor
influence on the values of relative alignments $\delta{I}$.

The presence of the second term studied here,
$\bbox{s}_t$$\cdot$$(\bbox{\nabla}$$\times$$\bbox{j}_t)$, has
very little effect on the values and relative values of ${{\cal J}^{(2)}}$.
However, its impact on the values of relative alignments is very
large.  Removing this term from the functional, decreases the
values of $\delta{I}$ by about 0.5\,$\hbar$. For the
$^{151}$Tb(2)--$^{152}$Dy(1)
pair of bands, this leads to relative alignment which has the
experimental value at the high-spin end of the band,
Fig.~\ref{fig09}(c). At the low-spin end the value of $\delta{I}$
is now underestimated by about 0.3\,$\hbar$. According to
Eq.~(\ref{eq216}), this gives the differences of the $\gamma$-ray
energies increasing from about $-$6\,keV to 0, i.e., the
identity of $E_\gamma$ is still not reproduced with a
sufficient precision.  The sensitivity of results to the
$C_t^{\nabla j}$ coupling constant suggests, however, that a fit
of this kind of term to experimental data may, in principle,
improve the agreement. As mentioned previously, such a fit
should take into account many different bands and cannot be
based on a rather restricted set of examples discussed here.

{}For the $^{150}$Gd(2)--$^{151}$Tb(1) pair of bands, the
removal of the
$\bbox{s}_t$$\cdot$$(\bbox{\nabla}$$\times$$\bbox{j}_t)$ term,
while keeping the term $\bbox{j}_t^2$ unchanged, leads to a very
good description of the values of ${{\cal J}^{(2)}}$, and of the relative
alignments $\delta{I}$ simultaneously, Fig.~\ref{fig10}. In this
case, the relative alignments agree with the experiment to
better than 0.1\,$\hbar$, which corresponds to the identity of
the $\gamma$-ray energies to better than 1\,keV.

\section{Conclusions}
\label{sec6}

In the present study we have applied the Hartree-Fock cranking
method with the Skyrme interactions to describe rotational
states in selected superdeformed nuclei. We have analyzed in detail
properties of two pairs of bands, namely, $^{151}$Tb(2)--$^{152}$Dy(1) and
$^{150}$Gd(2)--$^{151}$Tb(1). Experimentally, these bands are pair-wise
identical, i.e., the corresponding $\gamma$-transition energies
are identical within 2\,keV each.

In agreement with the previous interpretations,
we have fixed the configurations of the $^{151}$Tb(1) band as
a hole-structure created in the $\pi[651]{3/2}$~($r$=$+i$) or $\pi6_4$ Nilsson
intruder orbital with respect to the magic superdeformed
$^{152}_{~66}$Dy$_{86}$
core.  Similarly, the $^{151}$Tb(2) and $^{150}$Gd(2) bands have been
constructed as the $\pi[301]{1/2}$~($r$=$+i$) hole configurations in the
corresponding $^{152}$Dy(1) and $^{151}$Tb(1) cores.

The dynamical moments ${{\cal J}^{(2)}}$ calculated for the
$^{151}$Tb(2)--$^{152}$Dy(1) pair of bands and for the
$^{150}$Gd(2)--$^{151}$Tb(1)
pair of bands are identical within 2\,{$\hbar^2$MeV$^{-1}$}. This result
correctly reproduces the identity of these two pairs of bands
obtained in experiment. It confirms that the internal structure of
the $\pi[301]{1/2}$~($r$=$+i$) orbital is responsible for the
occurrence of these particular identical bands. The obtained identity of
${{\cal J}^{(2)}}$ does not depend on the version of the Skyrme force used,
neither it depends on including or disregarding various time-odd
components in the mean field of the rotating nucleus.  The fact
that the two bands calculated in $^{151}$Tb,
i.e., $^{151}$Tb(1) and $^{151}$Tb(2), are {\em not} identical
(similarly as in experiment), indicates a crucial role of the
single-particle structure of the involved orbitals, and shows
that in the HF cranking theory the phenomenon of identical bands
is not a generic, built-in-by-assumption result.

The relative alignments of the studied pairs of bands are more
difficult to reproduce than the simple identity of the
dynamical moments. First of all, the results do depend on the
version of Skyrme interaction and on the time-odd components
included in the mean field. When the complete Skyrme functionals
are used, i.e., when all time-odd terms are taken into account,
the calculated relative alignments do not reproduce the
experimental data. The disagreement obtained for the SkM*
interaction is particularly large, while the SIII and SkP
interactions also give too large relative alignments.  Within
the SkM* parametrization, good agreement with data is obtained
when all time-odd terms are disregarded except of the one which
involves the density of current, ${\bbox{j}}$, and is related to
the effective-mass term by the gauge transformation.

This specific result does not imply that the time-odd terms
present in the formalism should be generally eliminated from
its applications. By studying cases with some time-odd terms
removed we only aimed at gaining information which may
be useful when improving the parametrisation of effective interactions.

As seen from the results obtained in the present study, a
readjustment of parameters of effective forces in the time-odd
channel seems to be necessary for a detailed description of the
identical $\gamma$-ray transitions in superdeformed nuclei.
However, this should be done by taking into account many
available high-spin data simultaneously, and also should involve
a readjustment of the time-even part
(the spin-orbit interaction, in particular)
which is responsible for the ordering of the single-particle
energies and routhians. The work along these lines is now in progress.

\acknowledgments

We would like to express our thanks to the
{\it Institut du D\'eveloppement et de Ressources en Informatique
Scientifique} (IDRIS) of CNRS, France, which provided us with
the computing facilities under Project no.~940333.
This research was supported in part by the Polish Committee for
Scientific Research under Contract No.~2~P03B~034~08.

\appendix

\section{Coupling constants of the Skyrme functional}
\label{appA}

In its standard form, the Skyrme interaction (see
e.g.{} Ref.{} \cite{BON87a}) depends on 10 parameters,
$t_0$, $x_0$, $t_1$, $x_1$, $t_2$, $x_2$, $t_3$, $x_3$,
$\alpha$, and $W$. For $x_1$=$x_2$=0, the energy density which
corresponds to the Skyrme interaction has been derived in
Ref.{} \cite{EBG75}.  In terms of the isovector and isoscalar
coupling constants, and for the complete interaction, this energy
density is given in Eqs.~(\ref{eq108})--(\ref{eq109}).

The zero-order coupling constants, $C_t^{\rho}$ and $C_t^{
s}$, correspond to the velocity-independent terms of the
interaction, and can be expressed by the $t_0$, $x_0$, $t_3$,
and $x_3$ parameters, as it shows Table~\ref{tab2}.  The
zero-order time-odd coupling constants $C_t^{   s}$ (for $t$=0
and 1) are linear combinations of the zero-order time-even
coupling constants $C_t^{\rho}$ (Table~\ref{tab4}).

Similarly, the second-order coupling constants,
$C_t^{\Delta\rho}$, $C_t^{\tau}$, $C_t^{\Delta s}$, and
$C_t^{T}$, correspond to the velocity-dependent terms of the
interaction given by the parameters $t_1$, $x_1$, $t_2$, and
$x_2$, and are presented in Table~\ref{tab3}.  The second-order
time-odd coupling constants, $C_t^{\Delta s}$ and $C_t^{   T}$,
are linear combinations of the second-order time-even coupling
constants, $C_t^{\Delta\rho}$ and $C_t^{\tau}$ (Table~\ref{tab5}).
Another four second-order coupling constants, $C_t^{   J}$ and
$C_t^{   j}$, which also depend on the same $t_1$, $x_1$, $t_2$,
and $x_2$ parameters, are given by the gauge-invariance
conditions (\ref{eq215a}) and (\ref{eq215b}).

Finally, the second-order coupling constants $C_t^{\nabla J}$
are given by the spin-orbit term of the Skyrme interaction and
depend on the parameter $W$, $C_0^{\nabla J}$=$-$$\frac{3}{4}W$
and $C_1^{\nabla J}$=$-$$\frac{1}{4}W$, while the other two,
$C_t^{\nabla j}$, follow from the gauge-invariance
condition (\ref{eq215c}).  An extension of the Skyrme energy
density, which introduces the coupling constants $C_0^{\nabla
J}$ and $C_1^{\nabla J}$ independent of one another, is
discussed in Ref.{} \cite{Rei94}

\renewcommand{\topfraction}{0.0}
\renewcommand{\bottomfraction}{0.0}
\renewcommand{\floatpagefraction}{0.0}
\setcounter{topnumber}{20}
\setcounter{bottomnumber}{20}
\setcounter{totalnumber}{20}

\clearpage

\mediumtext
\begin{table}
\caption[T1]{Coupling constants in the gauge-invariant energy
density (\protect\ref{eq109c}) obtained for the SIII, SkP, and
SkM* Skyrme interactions. Zero-order coupling constants
$C_t^{\rho}$ and $C_t^{   s}$ ar given in {MeV$\cdot$fm$^3$} and the
remaining (second-order) ones are given in {MeV$\cdot$fm$^5$}.}
\label{tab1}
\begin{tabular}{l|rrr@{\hspace{0.5em}}|rrr}
\hline
&\multicolumn{3}{c@{\hspace{0.5em}}|}{$t$=0}&\multicolumn{3}{c}{$t$=1} \\
&  \multicolumn{3}{c|}{\hspace{-3em}\hrulefill}
&  \multicolumn{3}{c} {\hspace{-3em}\hrulefill} \\
& \text{SIII} & \text{SkM*} & \text{SkP}~
& \text{SIII} & \text{SkM*} & \text{SkP}~  \\
\hline
 $C_t^{\rho}(\rho\mbox{$=$}0)$
        & $-$423.3 & $-$991.9 & $-$1099.4 &    268.8 &    390.1 &    580.6 \\
 $C_t^{\rho}(\rho\mbox{$=$}\rho_{\text{nm}})$
        & $-$296.4 & $-$237.7 &  $-$335.6 &    141.2 &    150.8 &    188.4 \\
 $C_t^{\Delta\rho}$
        &  $-$63.0 &  $-$68.2 &   $-$60.1 &     17.0 &     17.1 &     35.1 \\
 $C_t^{\tau}$
        &     44.4 &     34.7 &       0.0 &  $-$30.6 &  $-$34.1 &  $-$44.6 \\
 $C_t^{\nabla J}$
        &  $-$90.0 &  $-$97.5 &   $-$75.0 &  $-$30.0 &  $-$32.5 &  $-$25.0 \\
\hline
 $C_t^{   s}(\rho\mbox{$=$}0)$
        &     14.1 &    271.1 &     152.3 &    141.1 &    330.6 &    366.5 \\
 $C_t^{   s}(\rho\mbox{$=$}\rho_{\text{nm}})$
        &     56.4 &     31.7 &   $-$31.4 &     98.8 &     91.2 &     78.5 \\
 $C_t^{\Delta s}$
        &     17.0 &     17.1 &    $-$4.2 &     17.0 &     17.1 &      9.8 \\
 $C_t^{   T}$
        &  $-$30.6 &  $-$34.1 &       7.7 &  $-$30.6 &  $-$34.1 &  $-$41.1 \\
\hline
\end{tabular}
\end{table}
\narrowtext\noindent%

\begin{table}
\caption[T6]{Proton quadrupole moments (in barns) calculated for
             the SIII, SkM*, and SkP Skyrme forces at
             $\omega$=0.5\,{MeV/$\hbar$}.}
\label{tab6}
\begin{tabular}{r@{}l|rrr}
\hline
   &     &   SIII   &   SkM*   &   SkP    \\
\hline
$^{152}$Dy&(1)  &  18.55   &  18.25   &  18.36   \\
$^{151}$Tb&(2)  &  18.69   &  18.38   &  18.48   \\
$^{151}$Tb&(1)  &  17.59   &  17.22   &  17.41   \\
$^{150}$Gd&(2)  &  17.74   &  17.36   &  17.54   \\
\hline
\end{tabular}
\end{table}

\begin{table}
\caption[T2]{Zero-order coupling constants as functions of
             parameters of the Skyrme interactions, expressed by
             the formula:  $C$=$\frac{1}{8}$($at_0$+$bt_0x_0$)+$
             \frac{1}{48}\rho^\alpha ($$at_3$+$bt_3x_3$).}
\label{tab2}
\begin{tabular}{l|rr}
\hline
& $a$ & $b$ \\
\hline
 $C_0^{\rho}$       &     3  &     0  \\
 $C_1^{\rho}$       &  $-$1  &  $-$2  \\
\hline
 $C_0^{   s}$       &  $-$1  &     2  \\
 $C_1^{   s}$       &  $-$1  &     0  \\
\hline
\end{tabular}
\end{table}

\begin{table}
\caption[T4]{Zero-order time-odd coupling constants as functions
             of the time-even coupling constants, expressed by
             the formula:
             $C$=$\frac{1}{3}$($aC_0^{\rho}$+$bC_1^{\rho}$).}
\label{tab4}
\begin{tabular}{l|rr}
\hline
& $a$ & $b$ \\
\hline
 $C_0^{   s}$       &  $-$2  &  $-$3  \\
 $C_1^{   s}$       &  $-$1  &     0  \\
\hline
\end{tabular}
\end{table}

\begin{table}
\caption[T3]{Second-order coupling constants as functions of
             parameters of the Skyrme interactions, expressed by
             the formula:  $C$=$\frac{1}{64}$($at_1$+$bt_1x_1$+$
             ct_2$+$dt_2x_2$).}
\label{tab3}
\begin{tabular}{l|rrrr}
\hline
& $a$ & $b$ & $c$ & $d$ \\
\hline
 $C_0^{\Delta\rho}$ &  $-$9  &     0  &     5  &     4   \\
 $C_1^{\Delta\rho}$ &     3  &     6  &     1  &     2   \\
 $C_0^{\tau}      $ &    12  &     0  &    20  &    16   \\
 $C_1^{\tau}      $ &  $-$4  &  $-$8  &     4  &     8   \\
\hline
 $C_0^{\Delta s}  $ &     3  &  $-$6  &     1  &     2   \\
 $C_1^{\Delta s}  $ &     3  &     0  &     1  &     0   \\
 $C_0^{   T}      $ &  $-$4  &     8  &     4  &     8   \\
 $C_1^{   T}      $ &  $-$4  &     0  &     4  &     0   \\
\hline
\end{tabular}
\end{table}

\begin{table}
\caption[T5]{Second-order time-odd coupling constants as
             functions of the time-even coupling constants,
             expressed by the formula:
             $C$=$\frac{1}{24}$($aC_0^{\Delta\rho}$+$bC_1^{\Delta\rho}$+$
                                 cC_0^{\tau}$+$dC_1^{\tau}$).}
\label{tab5}
\begin{tabular}{l|rrrr}
\hline
& $a$ & $b$ & $c$ & $d$ \\
\hline
 $C_0^{\Delta s}$   &     0  &     6  &     3  &     9   \\
 $C_1^{\Delta s}$   &  $-$4  &  $-$4  &     3  &  $-$3   \\
 $C_0^{   T}    $   &     0  &    48  &  $-$4  &    12   \\
 $C_1^{   T}    $   &    16  & $-$16  &     4  & $-$12   \\
\hline
\end{tabular}
\end{table}

\clearpage

\begin{figure}[ht]
\caption[F5]{%
Calculated dynamical moment ${{\cal J}^{(2)}}$ for the yrast
band of $^{152}$Dy, part (a), the relative dynamical moment
$\delta{{\cal J}^{(2)}}$ calculated for the $^{151}$Tb(2) and
$^{152}$Dy(1) bands, part (b),
and the relative alignment $\delta I$ between these two bands,
part (c). The experimental points are denoted by asterisks.
Complete Skyrme functionals of SIII, SkM*, and SkP interactions
have been used.  Note the scale in (b) expanded five times as
compared to (a).}
\label{fig05}
\end{figure}

\begin{figure}[ht]
\caption[F6]{%
Same as Fig.~\protect\ref{fig05}, but for the $^{151}$Tb(1) band, (a),
and for the differences between the $^{150}$Gd(2) and
$^{151}$Tb(1) bands, (b)
and (c).}
\label{fig06}
\end{figure}

\begin{figure}[ht]
\caption[F7]{%
Same as Fig.~\protect\ref{fig05}, but for the complete SkM*
functional (open squares) compared to three versions of the
modified SkM* functional, namely, $C_t^{   T}$=0 (open circles),
$C_t^{   T}$=$C_t^{\Delta s}$=0 (full squares), and
$C_t^{T}$=$C_t^{\Delta s}$=$C_t^{   s}$=0 (full circles).  The
modified functionals are gauge invariant, i.e., $C_t^{   J}$=0
whenever $C_t^{   T}$=0, Eq.(\protect\ref{eq215b}).}
\label{fig07}
\end{figure}

\begin{figure}[ht]
\caption[F8]{%
Same as Fig.~\protect\ref{fig07}, but for the
$^{150}$Gd(2)--$^{151}$Tb(1) pair
of bands.}
\label{fig08}
\end{figure}

\begin{figure}[ht]
\caption[F9]{%
Same as Fig.~\protect\ref{fig07}, but for the modified
gauge-invariant SkM* functional given by $C_t^{T}$=$C_t^{\Delta
s}$=$C_t^{   s}$=0 (full circles), compared to three versions of
the gauge-violating SkM* functional obtained by putting in
addition $C_t^{   j}$=0 (open symbols) and/or $C_t^{\nabla j}$=0
(squares). The version with $C_t^{j}$=$C_t^{\nabla j}$=0 (open
squares) corresponds to all mean-field time-odd terms neglected.
}
\label{fig09}
\end{figure}

\begin{figure}[ht]
\caption[F10]{%
Same as Fig.~\protect\ref{fig09}, but for the
$^{150}$Gd(2)--$^{151}$Tb(1) pair
of bands.
}
\label{fig10}
\end{figure}
%
% Upper-case    A B C D E F G H I J K L M N O P Q R S T U V W X Y Z
% Lower-case    a b c d e f g h i j k l m n o p q r s t u v w x y z
% Digits        0 1 2 3 4 5 6 7 8 9
% Exclamation   !           Double quote "          Hash (number) #
% Dollar        $           Percent      %          Ampersand     &
% Acute accent  '           Left paren   (          Right paren   )
% Asterisk      *           Plus         +          Comma         ,
% Minus         -           Point        .          Solidus       /
% Colon         :           Semicolon    ;          Less than     <
% Equals        =           Greater than >          Question mark ?
% At            @           Left bracket [          Backslash     \
% Right bracket ]           Circumflex   ^          Underscore    _
% Grave accent  `           Left brace   {          Vertical bar  |
% Right brace   }           Tilde        ~
%
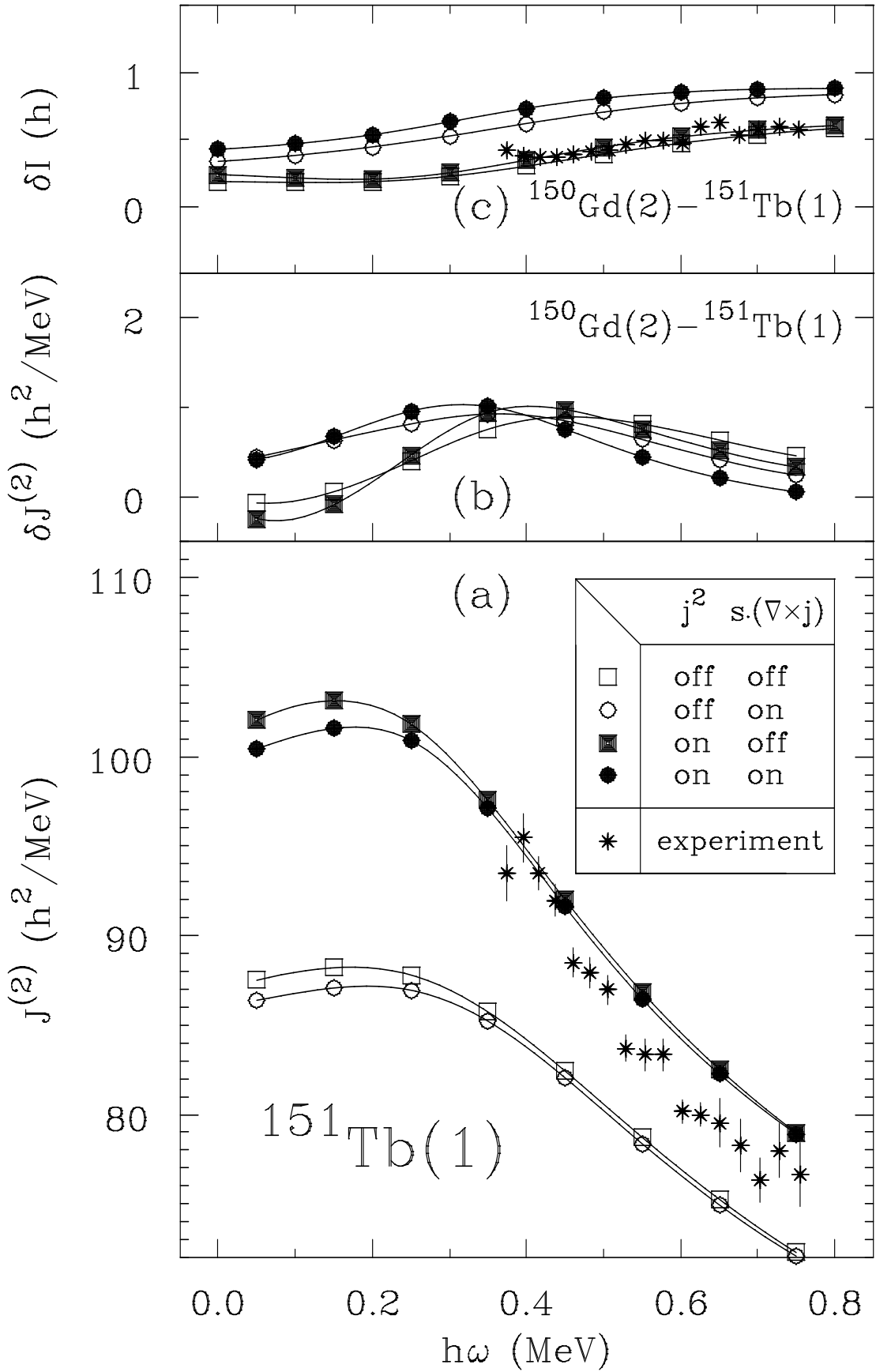
\end{document}